\documentclass[smallextended]{svjour3}

\smartqed  
\usepackage{graphicx}

\newtheorem{dfntn}{Definition}

\newtheorem{rmrk}{Remark}
\newcommand{\N}{{\rm I}\!{\rm N}}

\usepackage[colorlinks=true, a4paper=true, pdfstartview=FitV,
linkcolor=blue, citecolor=blue, urlcolor=blue]{hyperref}

\usepackage{amssymb}
\usepackage{amsmath}

\begin{document}

\title{Performability analysis of the second order semi-Markov chains: an application to wind energy production}

\author{Guglielmo D'Amico \and Filippo Petroni \and Flavio Prattico}

\institute{Dipartimento di Farmacia, Universit\`a G. d'Annunzio, 66013 Chieti, Italy\\
 \email{g.damico@unich.it} \and Dipartimento di Scienze Economiche e Aziendali, Universit\`a degli studi di Cagliari, 09123 Cagliari, Italy\\
 \email{fpetroni@unica.it}
\and Dipartimento di Ingegneria Meccanica, Energetica e Gestionale, Universit\`a degli studi dell'Aquila, 67100 L'Aquila, Italy\\
 \email{flavioprattico@gmail.com}}
 
 \date{Received: date / Accepted: date}
\maketitle

\begin{abstract}
In this paper a general second order semi-Markov reward model is presented.  Equations for the higher order moments of the reward process are presented for the first time and applied to wind energy production. The application is executed by considering a database, freely available from the web, that includes wind speed data taken from L.S.I. - Lastem station (Italy) and sampled every 10 minutes. We compute the expectation and the variance of the total energy produced by using the commercial blade Aircon HAWT - 10 kW. 
\end{abstract}



\section{Introduction}
Discrete time homogeneous semi-Markov chains have been recognized as a flexible and efficient tool in the modelling of stochastic systems. Recent results and applications are retrievable in \cite{barb04,jans07,dami12a}. The idea to link rewards to the occupancy of a semi-Markov state led to the construction of semi-Markov reward processes. These processes have been studied in \cite{howa71} and since then many developments and applications have been discussed. Non-homogeneous semi-Markov reward processes were defined in \cite{dedo86}. A method of computing the distribution of performability in a semi-Markov reward process was discussed in \cite{ciar90}. The asymptotic behaviour of a time homogeneous semi-Markov reward process was studied in \cite{solt98}.  More recent developments includes the derivations of higher order moments of the homogeneous semi-Markov reward process with initial backward (\cite{sten06}) and for the non-homogeneous case (\cite{sten07}), both the papers presented applications in the field of disability insurance. In  \cite{papa07} the reward paths in non-homogeneous semi-Markov systems in discrete time are examined with stochastic selection of the transition probabilities. The mean entrance probabilities and the mean rewards in the course of time are evaluated. This paper has been further generalized in \cite{papa12}. Finally in the paper by \cite{dami13b} duration dependent non-homogeneous semi-Markov chains were proposed in a disability insurance model.\\
\indent  In this paper we generalize some of the previous contributions by defining and analyzing the second order semi-Markov reward process in state and duration and giving relations for computing the higher order moments of this process. Moreover, we propose to employ a matrix notation that makes calculations easier and also provides a compact form for equations of moments of the reward process. The second order semi-Markov model in state and duration constitutes a generalization of the semi-Markov chain model because it allows the transition probabilities to vary depending on the last two visited states of the system and on the sojourn time lenght between these states.\\
\indent The second order semi-Markov model in state and duration was proposed in \cite{dami12} were it was applied in the modeling of wind speed. In that paper first and 
second order semi-Markov models were proposed with the aim of generating reliable synthetic wind speed data. There, it was shown that all the semi-Markov models perform better 
than the Markov chain model in reproducing the statistical properties of wind speed data. In particular, the model recognized as being more suitable is the second order semi-Markov 
chain in state and duration. This model has been further investigated in \cite{dami13} where classical reliability measures were computed with application 
to wind energy production. Continuing the effort in the searching of models ever more able to describe wind speed data, in this paper we apply the theoretical results concerning moments of the reward process to provide methods for computing the accumulated energy produced by a blade in a bounded time interval. The expected total energy produced gives important information on the feasibility of the investment in a wind farm and the riskiness of the investment can be measured in terms of variance, skewness, and kurtosis of the reward process. Finally notice that the technological characteristics of different blades are captured by the permanence reward and consequently we are able to choose among different blades to be installed at a given location.\\ 
\indent The paper is divided in this way: first, the second order semi-Markov model in state and duration is presented. Second, the reward structure is introduced and the equations 
of the higher order moments of the reward process are determined. Finally, the proposed approach is applied to compute moments of the total energy produced by a commercial blade 
applied to real wind speed data.

\section{The second order semi-Markov chain in state and duration}

In this section we give a short description of the second order semi-Markov chain in state and duration, see \cite{dami12} and \cite{dami13} for additional results.\\
\indent Let us consider a finite set of states $E=\{1,2,...,S\}$ in which the system can be into and a complete probability space $(\Omega, \emph{F}, \mathbb{P})$ on which we define the following random variables:
\begin{equation}
\label{uno}
J_{n}:\Omega\rightarrow E, \,\,\, T_{n}:\Omega\rightarrow \N .
\end{equation}
\indent They denote the state occupied at the $n$-th transition and the time of the $n$-th transition, respectively. To be more concrete, by $J_{n}$ we denote the wind speed at the $n$-th transition and by $T_{n}$ the time of the $n$-th transition of the wind speed.\\
\indent We assume that
\begin{equation}
\label{ventuno}
\begin{aligned}
& \mathbb{P}[J_{n+1}=j,T_{n+1}-T_{n}= t |\sigma(J_{s},T_{s}), J_{n}=k, J_{n-1}=i, T_{n}-T_{n-1}=x, 0\leq s \leq n]\\
& \quad =\mathbb{P}[J_{n+1}=j,T_{n+1}-T_{n}= t |J_{n}=k, J_{n-1}=i,T_{n}-T_{n-1}=x ]:=\,\,_{x}q_{i.k,j}(t).
\end{aligned}
\end{equation}
\indent Relation $(\ref{ventuno})$ asserts that, the knowledge of the values $J_{n}, J_{n-1},T_{n}-T_{n-1}$ suffices to give the conditional distribution of the couple $J_{n+1}, T_{n+1}-T_{n}$ whatever the values of the past variables might be. Therefore to make probabilistic forecasting we need the knowledge of the last two visited states and the duration time of the transition between them. For this reason we called this model a second order semi-Markov chains in state and duration.\\
\indent It should be remarked that in the paper by \cite{limn03} were defined nth order semi-Markov chains in continuous time. Anyway the dependence was only on past states and not on durations.\\ 
\indent The conditional probabilities 
$$
_{x}q_{i.k,j}(t)=\mathbb{P}[J_{n+1}=j, T_{n+1}-T_{n}=t|J_{n}=k,J_{n-1}=i,T_{n}-T_{n-1}=x ]
$$ 
are stored in a matrix of functions $\mathbf{q}=(_{x}q_{i.k,j}(t))$ named the second order kernel (in state and duration). The element $_{x}q_{i.k,j}(t)$ represents the probability that next wind speed will be in speed $j$ at time $t$ given that the current wind speed is $k$ and the previous wind speed state was $i$ and the duration in wind speed $i$ before of reaching wind speed $k$ was equal to $x$ units of time.\\
\indent From the knowledge of the kernel we can define the cumulated second order kernel probabilities:
\begin{equation}
\label{ventidue}
\begin{aligned}
& _{x}Q_{i.k,j}(t):=\mathbb{P}[J_{n+1}=j, T_{n+1}-T_{n}\leq t|J_{n}=k,J_{n-1}=i,T_{n}-T_{n-1}=x]\\
& =\sum_{s=1}^{t} {_{x}q_{i.k,j}(s)}.
\end{aligned}
\end{equation}
\indent The process $\{J_{n}\}$ is a second order Markov chain with state space $E$ and transition probability matrix $_{x}\mathbf{P}=\,_{x}\mathbf{Q}(\infty)$. We shall refer to it as the embedded Markov chain.\\
\indent Define the unconditional waiting time distribution function in states $k$ coming from state $i$ with duration $x$ as
\begin{equation}
\label{ventitre}
_{x}H_{i.k}(t):=\mathbb{P}[T_{n+1}-T_{n}\leq t|J_{n}=k,J_{n-1}=i,T_{n}-T_{n-1}=x]=\sum_{j\in E}{_{x}Q_{i.k,j}(t)}.
\end{equation}
\indent The conditional cumulative distribution functions of the waiting time in each state, given the state subsequently occupied is defined as
\begin{equation}
\label{ventiquattro}
\begin{aligned}
& _{x}G_{i.k,j}(t)=\mathbb{P}[T_{n+1}-T_{n}\leq t|J_{n}=k,J_{n-1}=i, J_{n+1}=j,T_{n}-T_{n-1}=x]\\
& =\frac{1}{_{x}p_{i.k,j}}\sum_{s=1}^{t}{_{x}q_{i.k,j}(s)}\cdot 1_{\{_{x}p_{i.k,j}\neq 0\}}+1_{\{_{x}p_{i.k,j}=0\}}
\end{aligned}
\end{equation}
\indent Define by $N(t)=\sup\{n\in \N:T_{n}\leq t\}$ $\forall t\in \N$. We define the second order (in state and duration) semi-Markov chain as $\mathbf{Z}(t)=(Z^{1}(t),Z^{2}(t))=(J_{N(t)-1},J_{N(t)}))$.\\
\indent For this model ordinary transition probability functions and transition probabilities with initial and final backward recurrence times were defined and computed in \cite{dami12} and reliability measures applied to wind energy production were presented in \cite{dami13}.

\section{The second order semi-Markov reward model in state and duration}
\label{Sdue}

In this section, following the line of research in \cite{sten07}, we determine recursive equations for higher order moments of the second order semi-Markov reward chain in state and duration.\\
\indent Let $\xi(t)$ denote the accumulated discounted reward during the time interval $(0,t]$ defined by the following relation,
\begin{equation}
\xi(t)=\sum_{0<u \leq t}\:  _{X_{N(u)-2}}\psi_{  J_{N(u)-2},J_{N(u)-1};J_{N(u)}} (B(u)) e^{-\delta u},
\label{1}
\end{equation}
where: $B(u)=u-T_{N(u)}$ is the backward recurrence time process, $ X_{N(u)}:=T_{N(u)+1}-T_{N(u)} $ is the sojourn time in state $J_{N(u)}$ before the $N(u)+1$ transition and $e^{-\delta}$ with $\delta \in [0,1]$ is the one period deterministic discount factor.\\
\indent The reward $_{X_{N(u)-2}}\psi_{  J_{N(u)-2},J_{N(u)-1};J_{N(u)}} (B(u))$ measures the performance of the system at time $u$. In our model the actual performance is a function of the current state $J_{N(u)}$ occupied by the system; it depends on the last two visited states $J_{N(u)-2}, J_{N(u)-1}$; it is duration dependent in the current state because it is a function of the backward process $B(u)$ and finally it depends on the sojourn time in the past states being dependent on $X_{N(u)-2}$. Our reward model is more general than those considered in \cite{sten07} because in that paper the authors considered a first order semi-Markov chain and the permanence reward $\psi$ do depend only on the couple $(J_{N(u)}, B(u))$.\\
\indent Let also denote by $ _{x,v}\xi_{i,k} (t) $ the random variable, which has the distribution the same with the conditional distribution for the random variable $\xi(t)$ given that 
$$J_{N(0)-2}=i,\: J_{N(0)-1}=k,  \: X_{N(0)-2}=x, \: B(T_{N(0)-1})=v,  $$
and let denote by $_{x,v}V_{i,k}^{(n)} (t):= E [( _{x,v}\xi_{i,k}(t))^n] $.\\
\indent In order to propose a matrix representation that simplifies calculations and provides a compact form for next equations we need to introduce the adopted notation and products.
\begin{dfntn}
Given two $m\times n$ matrices $\mathbf{A}=(A_{ij})$ and $\textbf{B}=(B_{ij})$, their Hadamard matrix product $\boxtimes$ gives the $m\times n$ matrix $C$ whose generic element is given by:
\[
C_{ij}=A_{ij}B_{ij}.
\]
\end{dfntn}
\begin{dfntn}
Let $\mathbf{A}$ be a $m^{2}\times m$ matrix and $\mathbf{B}$ be a $m^{2}\times 1$ vector, their $\otimes$ matrix product gives the $m^{2}\times 1$ vector whose elements, for all $i,k\in \{1,2,...,m\}$ are expressed by
\[
C_{(i-1)\cdot |m|+k}=\sum_{j=1}^{m}A_{(i-1)\cdot |m|+k,j}B_{(k-1)\cdot |m|+j,1}.
\] 
\end{dfntn}
\indent The first order moment of the reward process $ _{x,v}\xi_{i,k} (t) $ is computed in the following Theorem. 
\begin{theorem}
The first order moment of the second order semi-Markov chain in state and duration satisfies the following matrix equation:
\begin{equation}\label{fm}
\begin{aligned}
_{x, v }\mathbf{V}^{(1)}(t)=& _{x, v }\mathbf{D}(t) \boxtimes \: _{x, v }\mathbf{\tilde{\Psi}}(t) + \sum_{s=1}^{t}( _{x, v }\mathbf{B}(s) \cdot 1_{|E|}) \boxtimes \:  _{x, v }\mathbf{\tilde{\Psi}}(s)\\
& + \sum_{s=1}^{t} e^{-\delta s} \:  _{x, v }\mathbf{B}(s) \otimes \: _{x+s, 0 }\mathbf{V}^{(1)}(t-s) 
\end{aligned}
\end{equation} 
\noindent where $\forall i,k\in E$
\[
_{x,v}\mathbf{V}^{(1)}(t)=\left( _{x,v}V_{(i-1)\cdot |E|+k}^{(1)}(t)\right)=\left( _{x,v}V_{i,k}^{(1)}(t)\right),
\]
\[
_{x}\mathbf{\Psi}(v+u)= \left(  _{x}\psi_{(i-1)\cdot |E|+k}(v+u)e^{-\delta u}\right)=\left(  _{x}\psi_{i,k;k}(v+u)e^{-\delta u}\right),
\]
$_{x}\mathbf{\Psi}(v+u)$ is $|E|^2 \times 1$  and $_{x,v}\mathbf{\tilde{\Psi}}(t)=\sum_{u=1}^{t}\:_{x}\mathbf{\Psi}(v+u)$ is $|E|^2 \times 1$,
\[
_{x,v}\mathbf{D}(t)=\left( _{x,v}D_{(i-1)\cdot |E|+k}(t)\right)= \left( _{x,v}D_{i,k}(t) \right)
=\left(\frac{1-\, _{x}H_{i,k}(t+v)}{1-\, _{x}H_{i,k}(v)}\right)
\]
\[
_{x,v}\mathbf{B}(s)=\left( _{x,v}B_{(i-1)\cdot |E|+k,j}(s)\right)=\left( \frac{_{x}q_{i,k;j}(s+v)}{1-\, _{x}H_{i,k}(v)} \right)
\]
and ${\mathbf{1}_{|E|}}$ is the unitary row vector.
\end{theorem}
\textbf{Proof}:
Let consider the random variable $_{x,v}\xi_{i,k} (t)$. The time of next transition $T_{N(0)+1}$ can be greater of $t$ or not. Consequently it results that:
\begin{equation}
\label{somma}
_{x,v}V_{i,k}^{(1)}(t):=E [_{x,v}\xi_{i,k}(t)]=E [_{x,v}\xi_{i,k}(t)1_{\{T_{N(0)+1}>t\}}]+E [_{x,v}\xi_{i,k}(t)1_{\{T_{N(0)+1}\leq t\}}].
\end{equation}
\indent In the case $T_{N(0)+1}>t$ we have that
\begin{equation}
\label{2}
 _{x,v}\xi_{i,k} (t)= \sum_{u=1}^{t} \: _{x}\psi_{i,k;k} (u+v)e^{- \delta u}
\end{equation}
and this event occurs with probability 
\begin{equation}
\begin{aligned}
\label{tre}
& \mathbb{P}(T_{N(0)+1}>t | T_{N(0)+1}>0,  T_{N(0)}=-v, J_{N(0)}=k,  T_{N(0)-1}=-v-x, J_{N(0)-1}=i)\\
&=\frac{\mathbb{P}\left(  T_{N(0)+1}>t , T_{N(0)+1}>0,  T_{N(0)}=-v| J_{N(0)}=k,  T_{N(0)-1}=i,   X_{N(0)-1}=x \right)}{\mathbb{P}\left(   T_{N(0)+1}>0,  T_{N(0)}=-v| J_{N(0)}=k,  T_{N(0)-1}=i,  X_{N(0)-1}=x \right)}\\
&=\frac{\mathbb{P}\left(X_{N(0)}>t+v|J_{N(0)}=k,  T_{N(0)-1}=i,   X_{N(0)-1}=x \right)}{\mathbb{P}\left(X_{N(0)}>v|J_{N(0)}=k,  T_{N(0)-1}=i,   X_{N(0)-1}=x \right)}\\
& =\frac{1-\, _{x}H_{i,k}(t+v)}{1- \, _{x}H_{i,k}(v)}=\,_{x,v}D_{i,k}(t).
\end{aligned}
\end{equation}
\indent Then it results that
\begin{equation}
\label{dodici}
\mathbb{E} [_{x,v}\xi_{i,k}(t)1_{\{T_{N(0)+1}>t\}}]= \, _{x,v}D_{i,k}(t)\sum_{u=1}^{t} \: _{x}\psi_{i,k;k} (u+v)e^{- \delta u}.
\end{equation}
The elements $_{x,v}D_{i,k}(t)$ are stored in the matrix $_{x, v }\mathbf{D}(t)$ of dimension $|E|^{2}\times 1$ according to the following rule:
\[
\big(_{x, v }\mathbf{D}(t)\big)_{(i-1)\cdot |E|+k}:=_{x,v}D_{i,k}(t).
\]
\indent The elements $\sum_{u=1}^{t} \: _{x}\psi_{i,k;k} (u+v)e^{- \delta u}$ are stored in the matrix $_{x, v }\mathbf{\tilde{\Psi}}(t)$ of dimension $|E|^{2}\times 1$ according to the rule:
\[
\big(_{x, v }\mathbf{\tilde{\Psi}}(t)\big)_{(i-1)\cdot |E|+k}:=\sum_{u=1}^{t} \: _{x}\psi_{i,k;k} (u+v)e^{- \delta u}.
\]
\indent Consequently the right hand side of $(\ref{dodici})$ can be expressed in matrix form as follows:
\begin{equation}
\label{V1}
 _{x, v }\mathbf{D}(t) \boxtimes \: _{x, v }\mathbf{\tilde{\Psi}}(t).
\end{equation}
\indent In the second case, when $T_{N(0)+1} \leq t$, if we consider the next visited state $J_{N(0)+1}$ and the time of next transition $T_{N(0)+1}$ we have:
\begin{equation}
\label{exp2}
_{x,v}\xi_{i,k}(t) = \left(\sum_{s'=1}^{T_{N(0)+1}}\: _{x}\psi_{i,k;k}(v+s')e^{-\delta s'} + \: _{v+T_{N(0)+1},0}\xi_{k,J_{N(0)+1}}(t-s) e^{- \delta T_{N(0)+1}}\right).
\end{equation}
\indent The event $\{J_{N(0)+1}=j, T_{N(0)+1}=s\}$ occurs with probability 
\begin{equation}
\begin{aligned}
&\mathbb{P}\left( J_{N(0)+1}=j, T_{N(0)+1}=s|T_{N(0)+1}>0, T_{N(0)}=-v,J_{N(0)}=k,T_{N(0)-1}=-v-x,J_{N(0)-1}=i \right)\\
& =\frac{\mathbb{P}\left( J_{N(0)+1}=j, T_{N(0)+1}=s,T_{N(0)+1}>0, T_{N(0)}=-v|J_{N(0)}=k,T_{N(0)-1}=i,X_{N(0)-1}=x \right)}{\mathbb{P} \left(T_{N(0)+1}>0, T_{N(0)}=-v|J_{N(0)}=k,T_{N(0)-1}=i,X_{N(0)-1}=x \right)}\\
&=\frac{\mathbb{P}\left( J_{N(0)+1}=j, X_{N(0)}=s+v|J_{N(0)}=k,T_{N(0)-1}=i,X_{N(0)-1}=x \right)}{\mathbb{P} \left(X_{N(0)}>v|J_{N(0)}=k,T_{N(0)-1}=i,X_{N(0)-1}=x \right)}\\
&=\frac{_{x}q_{i,k;j}(s+v)}{1-\, _{x}H_{i,k}(v)}=\, _{x,v}B_{(i-1)\cdot |E|+k,j}(s).
\end{aligned}
\end{equation}
\indent Notice that the random variable $_{v+T_{N(0)+1},0}\xi_{k,J_{N(0)+1}}(t-s)$ is independent of the distribution of the joint random variable $(J_{N(0)+1}, T_{N(0)+1})$  because the accumulation process has the Markov property at transition times and consequently once the state  $J_{N(0)+1}$ and the $T_{N(0)+1}$ are known its behaviour doesn't depends on the distribution of $(J_{N(0)+1}, T_{N(0)+1})$. Then by taking the expectation in $(\ref{exp2})$ we get
\begin{equation}
\label{sedici}
\begin{aligned}
&\mathbb{E} [_{x,v}\xi_{i,k}(t)1_{\{T_{N(0)+1}\leq t\}}]\\ 
& =\sum_{j \in E} \sum_{s=1}^{t} \frac{_{x}q_{i,k;j}(s+v)}{1-\, _{x}H_{i,k}(v)} \cdot \sum_{s'=1}^{s}\: _{x}\psi_{i,k;k} (v+s')e^{- \delta s'}\\
&+ \sum_{j \in E} \sum_{s=1}^{t} \frac{_{x}q_{i,k;j}(s+v)}{1-\, _{x}H_{i,k}(v)} \cdot \, _{v+s,0}V_{k,j}^{(1)} (t-s)e^{-\delta s}.
\end{aligned}
\end{equation}
\indent The elements $\frac{_{x}q_{i,k;j}(s+v)}{1-\, _{x}H_{i,k}(v)}$ are stored in the matrix $_{x, v }\mathbf{B}(s)$ of dimension $|E|^{2}\times |E|$ according to the rule:
\[
\big( _{x, v }\mathbf{B}(s)\big)_{(i-1)\cdot |E|+k, j}:= \frac{_{x}q_{i,k;j}(s+v)}{1-\, _{x}H_{i,k}(v)}.
\]
\indent Then $\sum_{j \in E} \frac{_{x}q_{i,k;j}(s+v)}{1-\, _{x}H_{i,k}(v)}$ become the elements of the vector $_{x, v }\mathbf{B}(s) \cdot {\mathbf{1}_{|E|}}$ where  ${\mathbf{1}_{|E|}}$ is the unitary row vector.\\
\indent Having defined these matrices it is simple to realize that $( _{x, v }\mathbf{B}(s) \cdot {\mathbf{1}_{|E|}}) \boxtimes \:  _{x, v }\mathbf{\tilde{\Psi}}(s)$ is a $|E|^{2}\times 1$ vector and 
\[
\Big(( _{x, v }\mathbf{B}(s) \cdot {\mathbf{1}_{|E|}}) \boxtimes \:  _{x, v }\mathbf{\tilde{\Psi}}(s)\Big)_{(i-1)\times |E|+k}=\sum_{j \in E} \frac{_{x}q_{i,k;j}(s+v)}{1-\, _{x}H_{i,k}(v)} \cdot \sum_{s'=1}^{s}\: _{x}\psi_{i,k;k} (v+s')e^{- \delta s'}.
\]
\indent This argument, applied also to the second term on the r.h.s. of equation $(\ref{sedici})$, allows to represent $(\ref{sedici})$ in matrix form as:
\begin{equation}
\label{V2}
 \sum_{s=1}^{t}( _{x, v }\mathbf{B}(s) \cdot {\mathbf{1}_{|E|}}) \boxtimes \:  _{x, v }\mathbf{\tilde{\Psi}}(s) + \sum_{s=1}^{t} e^{-\delta s} \:  _{x, v }\mathbf{B}(s) \otimes \: _{x+s, 0 }\mathbf{V}^{(1)}(t-s).
\end{equation}
\indent A substitution of $(\ref{V1})$ and $(\ref{V2})$ in $(\ref{somma})$ concludes the proof.
\begin{flushright}
$\Box $
\end{flushright}
\indent By using similar techniques it is possible to get recursive equations for the higher order moments of the reward process.
\begin{corollary}
\indent The higher order moments of the reward process satisfy the following equation: 
\begin{equation}
\label{higher}
\begin{aligned}
&  _{x,v}V^{(n)}(t)=\, _{x,v}D(t) \boxtimes \, _{x,v}\tilde{\Psi}^{(n)}(t)+ \sum_{s=1}^{t} \left( _{x,v}B(s) \cdot 1_{|E|} \right) \boxtimes \, _{x,v}\tilde{\Psi}^{(n)}(s)\\
& + \sum_{s=1}^{t} e^{-\delta s n} \, _{x,v}B(s)\otimes \, _{v+s,0}V^{(n)}(t-s) \\
&+ \sum_{s=1}^{t}  \sum_{l=1}^{n-1}  \left( \begin{array}{c} n\\ l\\ \end{array} \right) \, _{x,v}\Psi^{(n-l)}(s) \boxtimes \left( e^{-\delta s l} \, _{x,v}B(s)\otimes \, _{v+s,0}V^{(l)}(t-s) \right).
\end{aligned}
\end{equation}
\end{corollary}
\textbf{Proof}:
\begin{equation}
\label{fund}
\begin{aligned}
_{x,v} V_{i,k}^{(n)}(t)&:= \mathbb{E}[\left( _{x,v}\xi_{i,k}(t) \right)^n]\\
& =E [(_{x,v}\xi_{i,k}(t))^{n}1_{\{T_{N(0)+1}>t\}}]+E [(_{x,v}\xi_{i,k}(t))^{n}1_{\{T_{N(0)+1}\leq t\}}].
\end{aligned}
\end{equation}
\indent In the case $T_{N(0)+1}>t$ we have that
\begin{equation}
( _{x,v}\xi_{i,k} (t))^{n}= \Big(\sum_{u=1}^{t} \: _{x}\psi_{i,k;k} (u+v)e^{- \delta u}\Big)^{n}
\end{equation}
and this event occurs with probability $_{x,v}D_{i,k}(t)$, see (\ref{tre}). Consequently it results that
\begin{equation}
\label{nuova}
E [_{x,v}\xi_{i,k}^{(n)}(t)1_{\{T_{N(0)+1}>t\}}]=\, _{x,v}D_{i,k}(t)\Big(\sum_{u=1}^{t} \: _{x}\psi_{i,k;k} (u+v)e^{- \delta u}\Big)^{n}.
\end{equation}
\indent In the opposite case, when $T_{N(0)+1}\leq t$, we have that
\begin{equation}
\begin{aligned}
\label{exp2n}
& _{x,v}\xi_{i,k}^{(n)}(t) = \left(\sum_{s'=1}^{T_{N(0)+1}}\: _{x}\psi_{i,k;k}(v+s')e^{-\delta s'} + \: _{v+T_{N(0)+1},0}\xi_{k,J_{N(0)+1}}(t-s) e^{- \delta T_{N(0)+1}}\right)^{n}\\
& =\Big(\sum_{s'=1}^{T_{N(0)+1}}\: _{x}\psi_{i,k;k}(v+s')e^{-\delta s'}\Big)^{n}+\, _{v+T_{N(0)+1},0}\xi_{k,J_{N(0)+1}}^{(n)}(t-s) e^{- n\delta T_{N(0)+1}}\\
& +\sum_{l=1}^{n-1} \left( \begin{array}{c} n\\ l\\ \end{array} \right) \left( \sum_{s'=1}^{s}\: _{x}\psi_{i,k;k} (v+s')e^{- \delta s'} \right)^{n-l}\Big( _{v+T_{N(0)+1},0}\xi_{k,J_{N(0)+1}}^{(l)}(t-s) e^{- l\delta T_{N(0)+1}}\Big).
\end{aligned}
\end{equation}
\indent The event $\{J_{N(0)+1}=j, T_{N(0)+1}=s\}$ occurs with probability $ _{x,v}B_{(i-1)\cdot |E|+k,j}(s)$.\\
\indent Then, by using the already mentioned independence between $_{v+T_{N(0)+1},0}\xi_{k,J_{N(0)+1}}(t-s)$ and the joint random variable $(J_{N(0)+1}, T_{N(0)+1})$ , by taking the expectation in $(\ref{exp2n})$ we get:
\begin{equation}
\begin{aligned}
\label{ultima}
& \mathbb{E}[ _{x,v}\xi_{i,k}^{(n)} (t)1_{\{T_{N(0)+1}\leq t\}}]= \\
&+ \sum_{j \in E} \sum_{s=1}^{t} \frac{_{x}q_{i,k;j}(s+v)}{1-\, _{x}H_{i,k}(v)} \cdot \left( \sum_{s'=1}^{s}\: _{x}\psi_{i,k;k} (v+s')e^{- \delta s'} \right)^n\\
&+ \sum_{j \in E} \sum_{s=1}^{t} \frac{_{x}q_{i,k;j}(s+v)}{1-\, _{x}H_{i,k}(v)} \cdot \, _{v+s,0}V_{k,j}^{(n)} (t-s)e^{-\delta s n}\\
& +\sum_{j \in E} \sum_{s=1}^{t} \sum_{l=1}^{n-1}  \frac{_{x}q_{i,k;j}(s+v)}{1-\, _{x}H_{i,k}(v)}  \left( \begin{array}{c} n\\ l\\ \end{array} \right) \left( \sum_{s'=1}^{s}\: _{x}\psi_{i,k;k} (v+s')e^{- \delta s'} \right)^{n-l}\\
& \cdot \, _{v+s,0}V_{k,j}^{(l)} (t-s)e^{-\delta s l}.
\end{aligned}
\end{equation}
\indent If we substitute $(\ref{nuova})$ and $(\ref{ultima})$ in $(\ref{fund})$ and we represent the resulting expression in matrix form we obtain the equation $(\ref{higher})$.
\begin{flushright}
$\Box $
\end{flushright}
\begin{rmrk}
If $\forall i\in E$ and $\forall x\in \N$ we have that
\[
_{x}q_{i.k,j}(t)=q_{k,j}(t),\,\,_{x}\psi_{i.k,j}(t)=\psi_{k,j}(t)
\]
then the second order semi-Markov reward chain model in state and duration collapses in a standard semi-Markov reward chain model and we recover exactly the results by \cite{sten07}.
\end{rmrk}
\begin{rmrk}
The adopted matrix notation for the moments of the second order semi-Markov reward chain in state and duration permits the computation of the moments with no more difficulties as compared to those necessary for the standard semi-Markov chain reward model.
\end{rmrk}

\section{Application to wind energy production}
In two previous papers \cite{dami12,dami13} we showed that wind speed can be well described by a second order semi-Markov process. Given these results we try to verify here if the reward model, described in the previous section, is able to well describe the production of energy by a wind turbine.
The data used in this analysis are freely available from $http://www.lsi-lastem.it/meteo/page/dwnldata.aspx$. 
The database is composed of about 230000 wind speed measures ranging from 0 to 16 $m/s$ with a sample frequency of 10 minutes. More accurate information about our database can be found in \cite{dami12,dami13}.

\begin{figure}
\centering
\includegraphics[height=6cm]{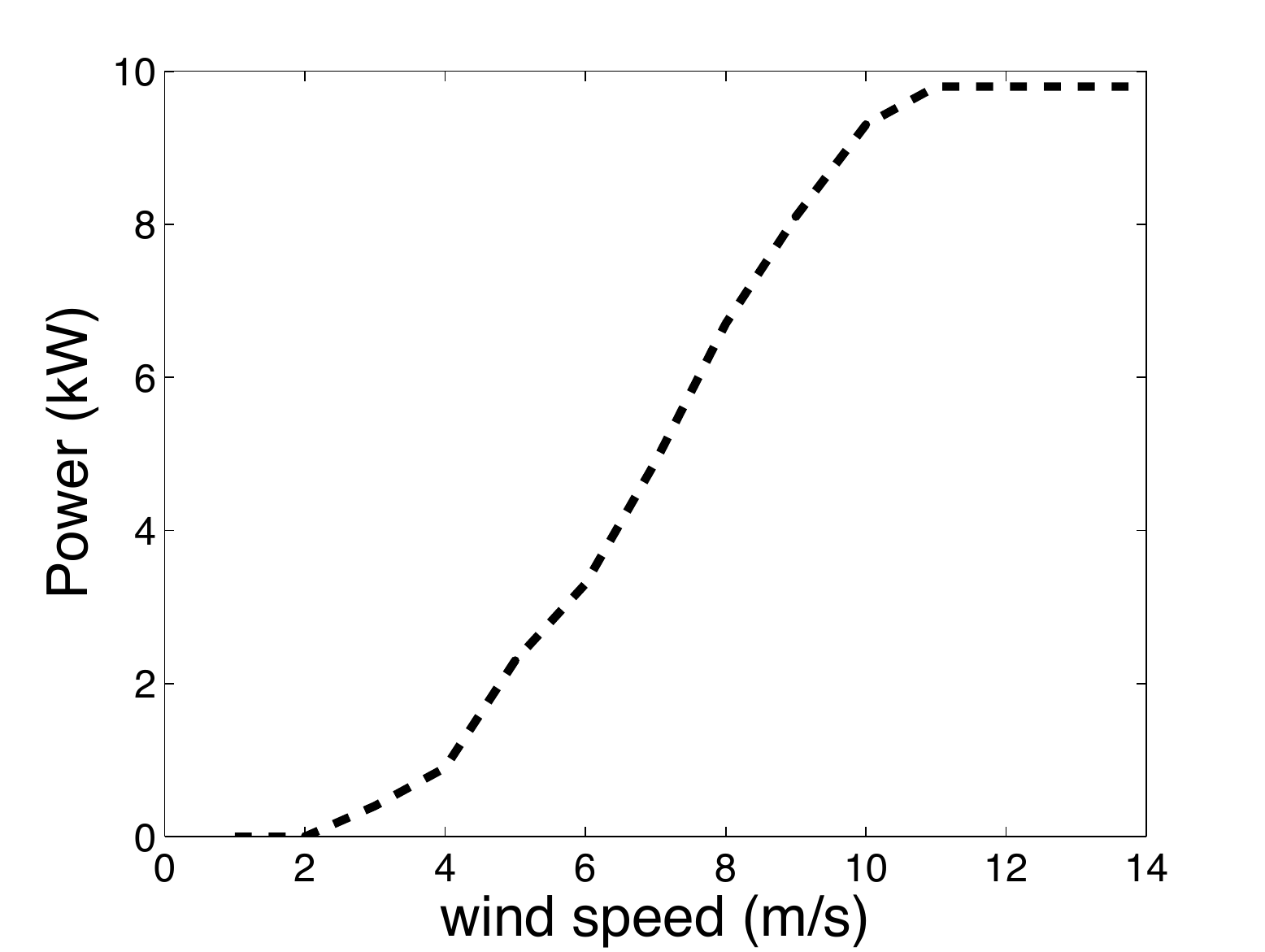}
\caption{Power curve of the 10 kW Aricon HAWT wind turbine.}\label{power}
\end{figure}

The state space of wind speed has been discretized into 8 states chosen to cover all the wind speed distribution. The state space is numerically represented by the set $E=\{0-1,1-2,2-3,3-4,4-5,5-6,6-7,>7\}(m/s)$. From the discretized trajectory $H(M)$ of the wind speed process 
\[
H(M)=\{J_{-1}, T_{-1}, J_{0}, T_{0}, ..., J_{N(M)}, u_{M}\},
\]
we have to estimate the probabilities ${\bf P}$ and ${\bf G}$. The quantity $u_{M}=M-T_{N(M)}$ is the censored sojourn time in the last wind speed state $J_{N(M)}$.  First of all we introduce the following counting processes:
\begin{equation}
\label{counting1}
_{x}N_{i.k,j}(t;M)=\sum_{n=1}^{N(M)}1_{\{J_{n}=j, J_{n-1}=k, J_{n-2}=i, X_{n-1}=x, X_{n}=t,\}}.
\end{equation}
\indent Formula $(\ref{counting1})$ expresses the number of transitions from the state $k$ to the state $j$ with a sojourn time $t$ which are preceded by a transition from the state $i$ into the state $k$ with sojourn time equal to $x$.\\
\begin{equation}
\label{counting2}
_{x}N_{i.k,j}(M)=\sum_{n=1}^{N(M)}1_{\{J_{n}=j, J_{n-1}=k, J_{n-2}=i, X_{n-1}=x\}}=\sum_{t>0}\,_{x}N_{i.k,j}(t;M).
\end{equation}
\indent Formula $(\ref{counting2})$ expresses the number of transitions from the state $k$ to the state $j$ which are preceded by a transition from the state $i$ into the state $k$ with sojourn time equal to $x$.\\
\begin{equation}
\label{counting3}
_{x}N_{i.k}(M)=\sum_{n=1}^{N(M)}1_{\{J_{n-1}=k, J_{n-2}=i, X_{n-1}=x\}}=\sum_{j\in E}\, _{x}N_{i.k,j}(M).
\end{equation}
\indent Formula $(\ref{counting3})$ expresses the number of transitions from the state $i$ into the state $k$ with sojourn time equal to $x$.\\
\indent The transition probabilities of the embedded Markov chain 
$$
_{x}p_{i.k,j}=\mathbb{P}[J_{n+1}=j|J_{n}=k,J_{n-1}=i,T_{n}-T_{n-1}=x ]
$$ 
can be estimated by
\begin{equation}
\label{stimap}
_{x}\hat{p}_{i.k,j}(M):=\frac{_{x}N_{i.k,j}(M)}{_{x}N_{i.k}(M)}.
\end{equation}
\indent The estimation of the conditional waiting time distributions is executed by considering the corresponding probability mass functions 
$$
_{x}g_{i.k,j}(t):=\,_{x}G_{i.k,j}(t)-\,_{x}G_{i.k,j}(t-1)
$$ 
which can be estimated by
\begin{equation}
\label{stimag}
_{x}\hat{g}_{i.k,j}(t;M):=\frac{_{x}N_{i.k,j}(t;M)}{_{x}N_{i.k,j}(M)}.
\end{equation}
\indent If $_{x}N_{i.k}(M)=0$ then $_{x}\hat{p}_{i.k,j}(M)=0$. If $_{x}N_{i.k,j}(M)=0$ then $_{x}\hat{g}_{i.k,j}(t;M)=0$.\\
\indent Starting from estimators $(\ref{stimap})$ and $(\ref{stimag})$ it is possible to obtain estimators of all the quantities of interest through a plug-in procedure.\\
\indent To have a realistic energy production we choose a commercial wind turbine, a 10 kW Aircon HAWT with a power curve given in Figure \ref{power}.
The power curve of a wind turbine represents how it produces energy as a function of the wind speed. In this case we have no production of energy in the interval 0-2 $m/s$, the wind turbine produces energy linearly from 3 $m/s$ to 10 $m/s$, then, with increasing wind speed the production remains constant until the limit of wind speed in which the wind turbine is stopped for structural reason. Note that the power curve is a graphical representation of the rewards $_{X_{N(u)-2}}\psi_{  J_{N(u)-2},J_{N(u)-1};J_{N(u)}} (B(u)) e^{-\delta u}$. In this application the rewards depend only on the present wind speed and $\delta$ is settled to be zero.

\begin{figure}
\centering
\includegraphics[height=4cm]{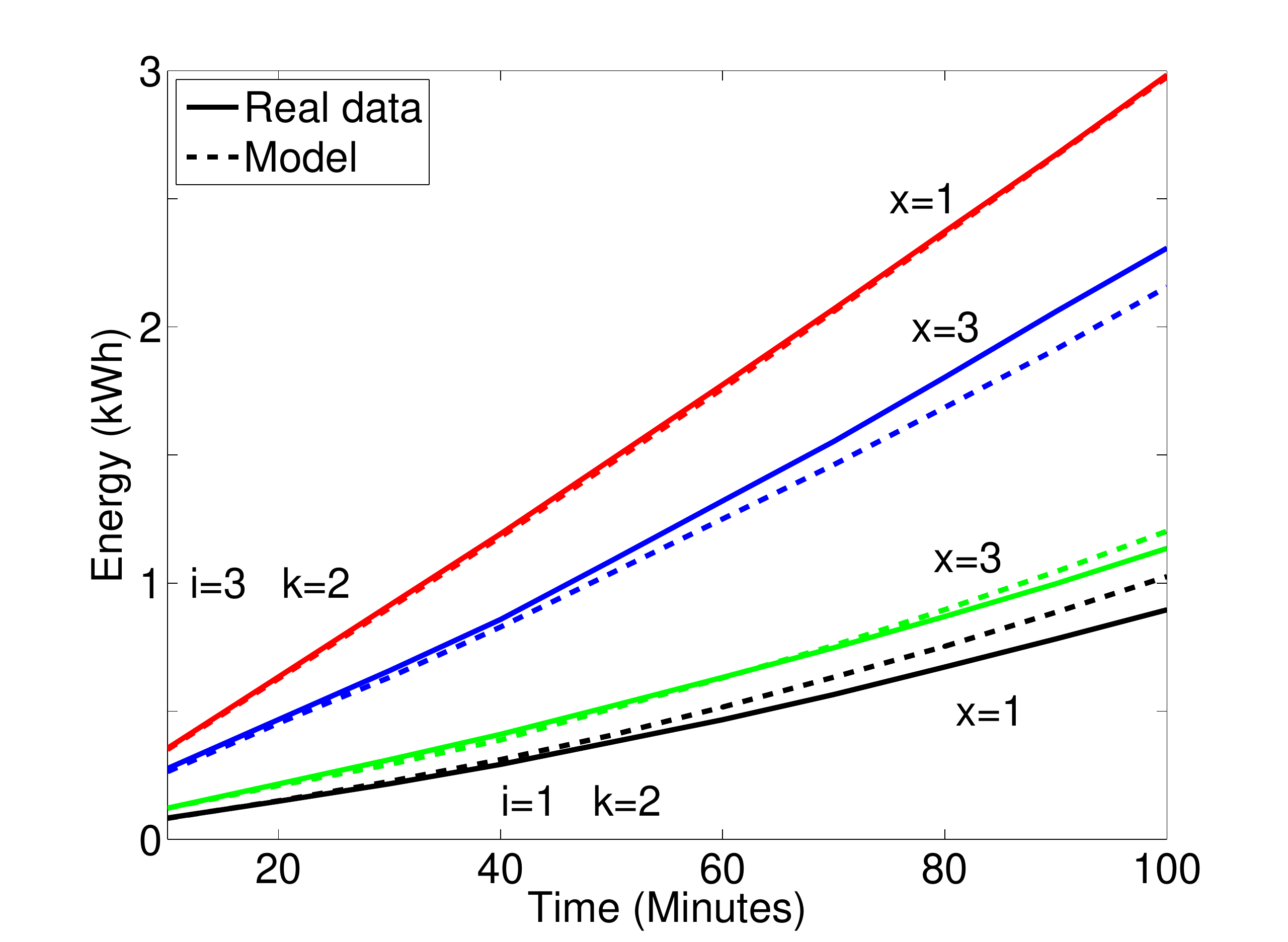}
\includegraphics[height=4cm]{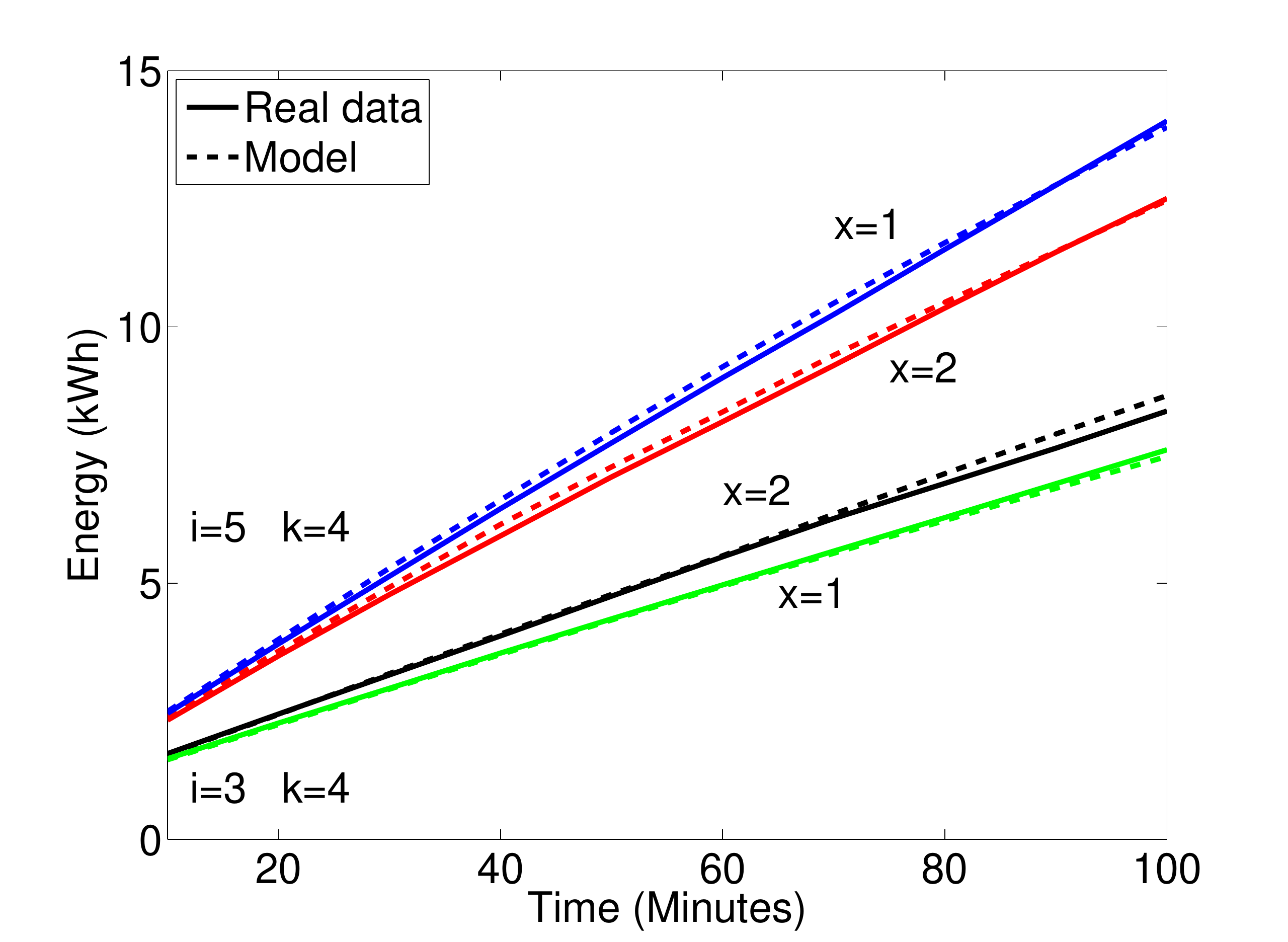}
\caption{Comparison of cumulated energy produced by real data and model when the current speed state is $k=2$ (left panel) and current speed state is $k=4$ (right panel)}\label{ener1}
\end{figure}

In Figure \ref{ener1} we compare the average cumulated energy produced by real data and the expected value $_{x}V^{(1)}_{i,k}(t)$ as calculated using formula $(\ref{fm})$ where $\delta=0$. The comparison is made by varying the sojourn time and the starting state. Particularly, the cumulated energy is plotted for two different initial states \textit{i} maintaining constant the current state \textit{k} for two different sojourn times \textit{x}. Left and right panel of the figure have different values for the current speed state. 

It is possible to note that all the cumulated energies plotted above depend strongly on the initial and current states and that there is also a great dependence on the sojourn time $x$. 
In fact, the expected value $_{x}V^{(1)}_{i,k}(t)$ has different values also if only the sojourn time $x$ is changed keeping constant initial states $i$ and final state $k$. For example, from Figure \ref{ener1} left panel it is possible to see that
$$
_{3}V^{(1)}_{1,2}(s) > \,_{1}V^{(1)}_{1,2}(s) \, \forall s\in [0,100],
$$
while
$$
_{3}V^{(1)}_{3,2}(s) <\,_{1}V^{(1)}_{3,2}(s) \, \forall s\in [0,100].
$$
This reveals that it is important to dispose of a model that is able to distinguish between these different situations which are determined only from a different duration of permanence in the initial state $i$ before making a transition to the current state $k$. Models based on Markov chains or classical semi-Markov chain are unable to capture this important effect that our second order semi-Markov chain in state and duration reproduces according to the real data.\\

\begin{figure}
\centering
\includegraphics[height=4cm]{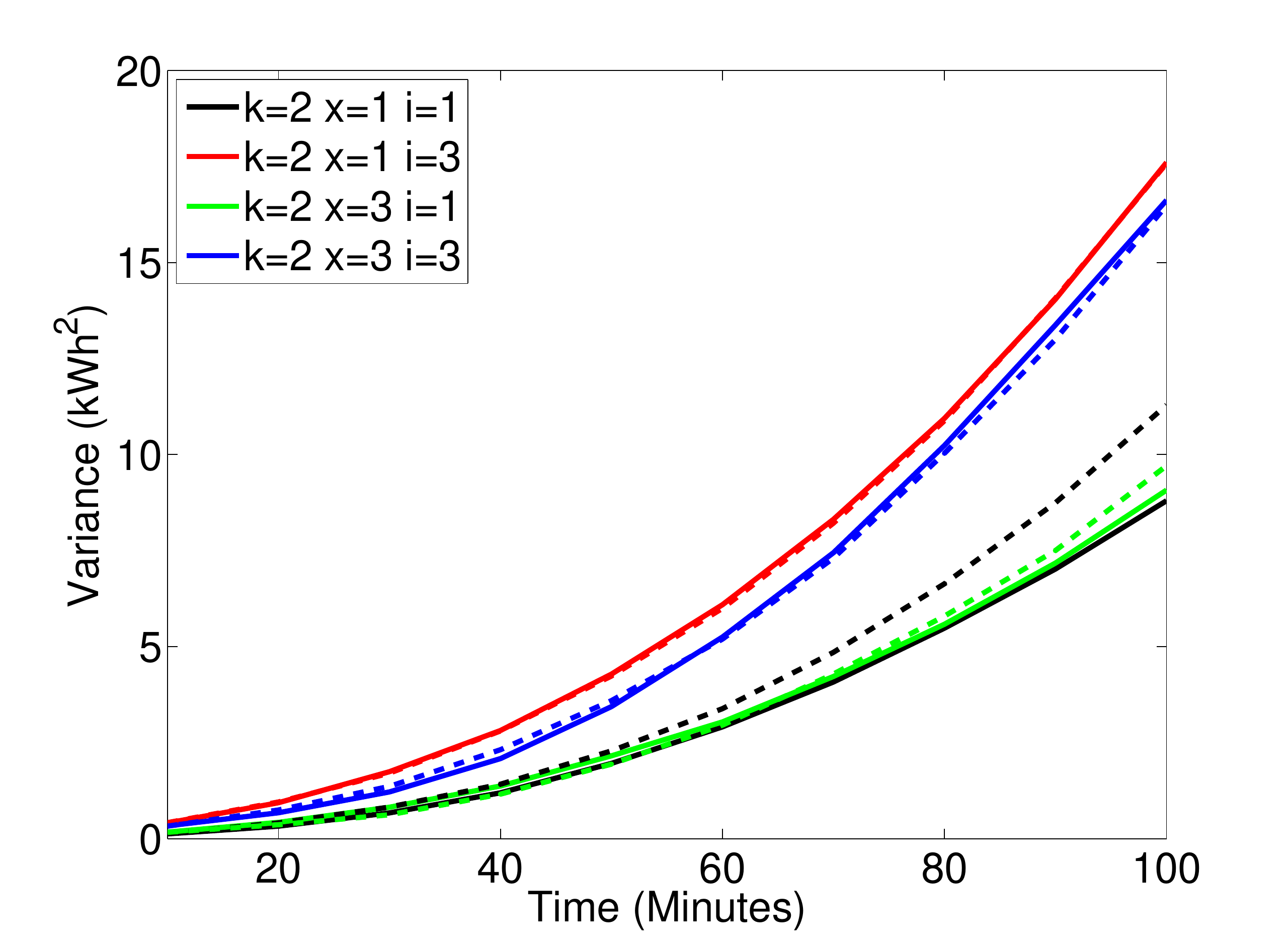}
\includegraphics[height=4cm]{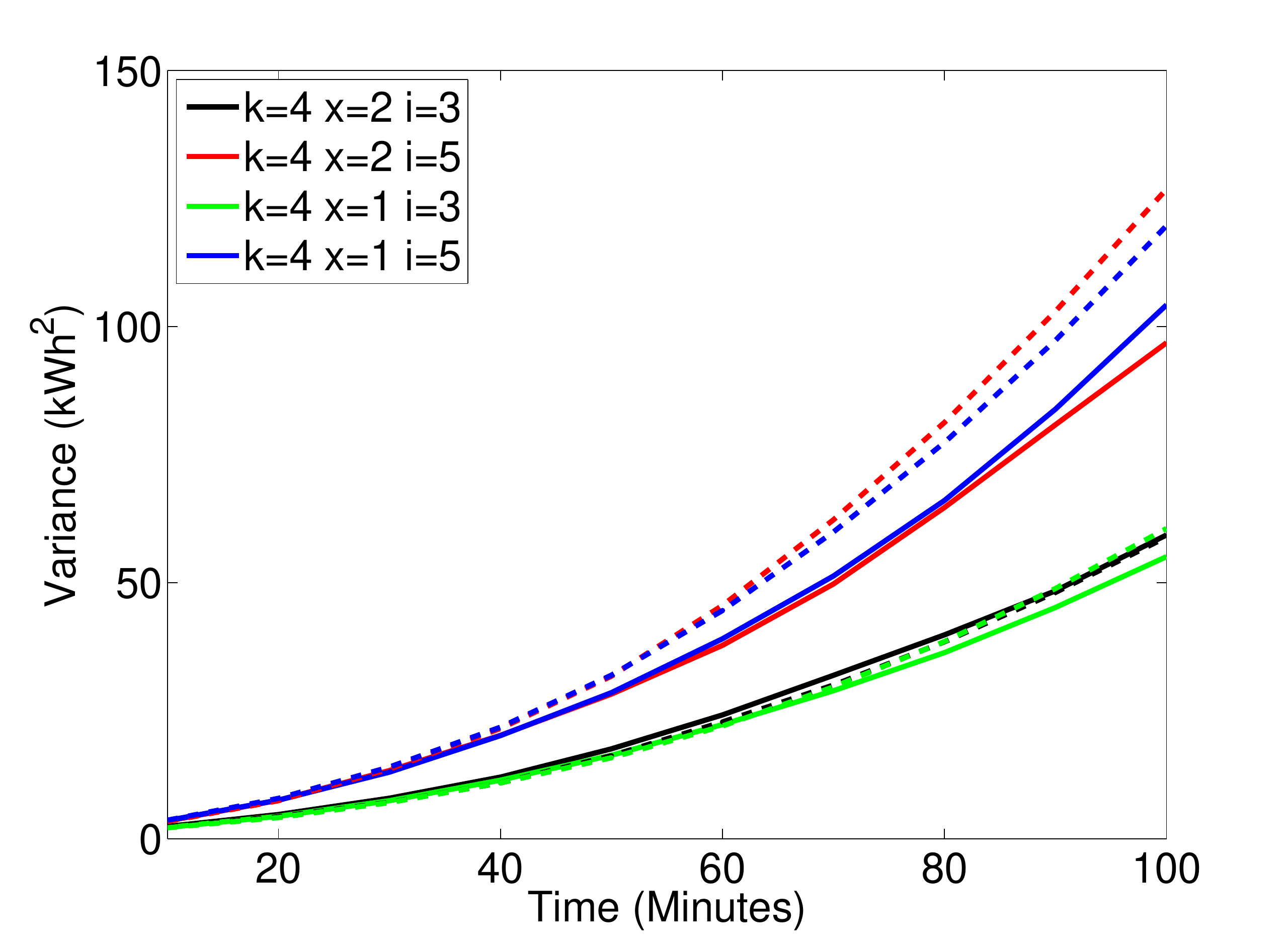}
\caption{Comparison of the variance of energy produced by real data (continuous line) and model (dashed line) when the current speed state is $k=2$ (left panel) and current speed state is $k=4$ (right panel)}\label{ener2}
\end{figure}

In Figure \ref{ener2} we compare the second central moment of the reward process again for real data and as obtained from the model calculated according to equation $(\ref{higher})$.
Also in this case it is possible to recognize that the model describes well real data behavior especially when the dependence from current states, past states and sojourn times is concerned. In this case the dependence is less evident but still present.

\section{Conclusion}
The purposes of this paper is to provide theoretical methods for computing the moments of a second order semi-Markov reward process in state and duration. The model is then used also to provide theoretical methods for computing the cumulated energy produced by a blade in a temporal interval $[0,T]$. 
We have reached this aim by modeling wind speed as a second order semi-Markov process. All the equations for the reward process are then derived under this hypothesis. Our model is tested against real data on wind speed freely available from the web.
We have shown that the proposed model is able to reproduce well the behavior of real data as far as energy production from wind speed is concerned. In particular, we have shown that the cumulated energy produced by a commercial blade does depend on the initial state \textit{i}, the current state \textit{k} and on the sojourn times \textit{x}. These results confirm the second order semi-Markov hypothesis.

\end{document}